%% file: main.tex
\documentclass[utf8]{FrontiersinHarvard} %
\usepackage{url,hyperref,lineno,microtype,subcaption}
\usepackage{graphicx,wrapfig,lipsum,epstopdf}
\usepackage{array}
\usepackage{amssymb}
\usepackage{tikz}
\usepackage{amsmath, amsfonts}
\usepackage[utf8]{inputenc}
\usepackage{listings}
\usepackage{mathrsfs}
\usepackage{xcolor}
\usepackage[colorinlistoftodos]{todonotes}
\usepackage{caption}
\usepackage{subcaption}
\usepackage[T1]{fontenc}
\usepackage{xcolor}
\usepackage{siunitx}
\usepackage{braket}
\usepackage{physics}
\usepackage{multicol}
\usepackage{algpseudocode}
\usepackage{algorithm}
\usepackage{program}
\usepackage{stackengine}
\usepackage{bookmark}
\usepackage{amsmath,esint}
\usepackage{adjustbox}
\usepackage{fancyhdr}
\newcommand*\dif{\mathop{}\!\mathrm{d}}
\usepackage{fancyhdr}
\numberwithin{equation}{section}
\makeatletter
\newcommand\footnoteref[1]{\protected@xdef\@thefnmark{\ref{#1}}\@footnotemark}

\usepackage[english,ngerman]{babel}

\def\keyFont{\fontsize{8}{11}\helveticabold }
\def\firstAuthorLast{Hansen {et~al.}} 
\def\Authors{Mikkel T. Hansen\,$^{1,*}$, Emil Z. Ulsig\,$^{1}$, Fabien Labbé\,$^{2}$, Magnus L. Madsen\,$^{1}$, Yunhong Ding\,$^{2}$, Karsten Rottwitt\,$^{2}$, and Nicolas Volet\,$^{1}$ }

\def\Address{$^{1}$ Department of Electrical and Computer Engineering, Aarhus University, Aarhus, Denmark \\
$^{2}$ DTU Electro, Department of Electrical and Photonics Engineering, Technical University of Denmark, Kongens Lyngby, Denmark }
\def\corrAuthor{Mikkel T. Hansen}
\def\corrEmail{torrild@ece.au.dk}

\begin{document}
\selectlanguage{english}

\onecolumn
\firstpage{1}

\title {Efficient and robust 
second-harmonic generation 
in thin-film lithium niobate using modal phase matching} 

\author[\firstAuthorLast ]{\Authors} 
\address{} 
\correspondance{} 

\extraAuth{}

\maketitle


\begin{abstract}
\section{}
A double-ridge waveguide 
is designed
for efficient and 
robust second-harmonic generation (SHG)
using 
the thin-film
lithium-niobate-on-insulator (LNOI)
platform.
Perfect phase matching (PhM) is achieved 
between the fundamental waveguide mode 
at 1550 nm
and a higher-order mode at the second harmonic.
The fabrication tolerances of the PhM condition are simulated using 
a finite-difference method mode solver, 
and conversion efficiencies as high as  
3.92 W$^{-1}$ are obtained
for a 1-cm long
waveguide.
This design allows access to the largest 
element of the second-order nonlinear susceptibility tensor, 
and represents 
a scalable alternative to waveguides based on 
periodically-poled lithium niobate (PPLN). The design has the potential for generating pairs of entangled
photons in the infrared C-band by spontaneous parametric down-conversion (SPDC).


\tiny
 \keyFont{ \section{Keywords:} thin-film LNOI, frequency conversion, second harmonic generation, single photons, higher order modes, nonlinear waveguides.} 
\end{abstract}

\input{Sections/Introduction}

\input{Sections/design}

\input{Sections/method}

\input{Sections/Results}
\input{Sections/Discussion}
\input{Sections/ConclusionOutlook}

\bibliographystyle{Frontiers-Harvard}

\bibliography{references}

\end{document}

%% file: Sections/Introduction.tex
\section{Introduction}
Single-photon sources (SPS) generating entangled photons are key components in various quantum devices 
with applications within quantum communication, computing, and metrology \citep{guo_parametric_2017}; \citep{obrien_photonic_2009};\citep{aharonovich_solid-state_2016}. 
Among the more established technologies are quantum encrypted communication systems \citep{basset_daylight_2023} such as quantum key distribution (QKD) protocols, which are based on entangled photons as messengers of secure information
\citep{zhang_device-independent_2022}.
SPS based on fluorescent semiconductor nanoparticles such as quantum dots (QD) constitute the prominent modality for generating single photons on-demand \citep{holewa_state---art_2023}. GaAs QD have shown to produce semi-deterministic photons of high indistinguishability and strong brightness \citep{huber_highly_2017}, while high-quality InAs QD have been interfaced with advanced photonic integrated circuits (PICs) \citep{wang_deterministic_2023-1}. QD-SPS are, however, in general challenged by inhomogeneous distribution of the nanoparticles \citep{aharonovich_solid-state_2016} and decoherence mechanism that results in linewidth broadening and consequently limited indistinguishability \citep{uppu_quantum-dot-based_2021}. An alternative and well-established method for generating highly entangled photons is the nonlinear frequency conversion process denoted spontaneous parametric down-conversion (SPDC).
SPDC is a second-order nonlinear interaction in which a photon interacts with a nonlinear medium such to generate two daughter photons at lower frequencies. This is a spontaneous process, hence SPS based on SPDC do not constitute deterministic sources in which the photon pair is generated on demand when triggered by a pump-pulse \citep{lodahl_deterministic_2022}. 
However, since the two generated photons are inherently
entangled, detecting one photon heralds the creation of the other. SPDC is, for this reason, typically referred to as a source of single \textit{heralded} photons. 

For the SPS to be feasible for practical quantum communications applications over long distances, it should preferably emit in the infrared C band, where low-loss fiber networks are already well-established. In this regard, lithium niobate (LN) has long been known for its unique optical properties, such as its high second-order nonlinearity and wide transparency window in the visible to infrared (VIS-IR) \citep{zhu_integrated_2021}. Though, recent advances in LN fabrication have renewed the interest in the material, where ultra-low losses in the thin-film (TF) LNOI platform has been reported \citep{zhang_monolithic_2017};\citep{shams-ansari_reduced_2022}. With the additional benefits of the strong confinement that can now  be achieved in TF-LNOI, the platform constitutes a powerful and scalable platform \citep{adcock_quantum_2022}, which is ideal for SPDC to the C-band. Typically, efficient frequency conversion in TF-LNOI is achieved by electro-optic modulation \citep{niu_research_2022};\citep{chen_ultra-broadband_2022};\citep{shah_visible-telecom_2023};\citep{hao_second-harmonic_2020};\citep{rao_actively-monitored_2019} such as periodically poled waveguides.  However, the fabrication of poled waveguides in TF-LNOI is complex \citep{chang_thin_2016}. To provide a scalable alternative to the method, this report considers modal phase matching (PhM), in which the geometrical parameters of the waveguide are adjusted to achieve efficient frequency conversion. Now, the conditions that govern efficient frequency conversion by SPDC also apply to the reversed classical conversion process, namely SHG \citep{helt_how_2012}. For this reason, the waveguide design is optimized for SHG in the simpler classical theoretical framework. A double-ridge waveguide is hence designed for efficient and
robust SHG at 775 nm based on the TF-LNOI platform. 
Perfect PhM is achieved 
between the fundamental TE waveguide mode 
at 1550-nm wavelength 
and a higher-order TE mode at the second harmonic, allowing access to the largest 
element of the second-order nonlinear 
susceptibility tensor. After thoroughly validating the numerical model, the theoretical maximal conversion efficiency is found to be comparable to the state-of-the-art efficiency in SHG. While similar studies of single-pass waveguide designs have been reported \citep{briggs_simultaneous_2021}, also where higher-order modes have been deployed \citep{luo_highly_2018};\cite{wang_second_2017}, the theoretical and simulations results from these display significantly lower conversion efficiencies than what is presented in this report. Additionally, this report addresses the fabrication tolerances of the design, identifies the critical parameters and provide a solution to a robust design.

%% file: Sections/design.tex
\section{Design} \label{sec:waveguidedesign}
\begin{figure}[htbp]
    \centering
    \includegraphics[width=\textwidth]{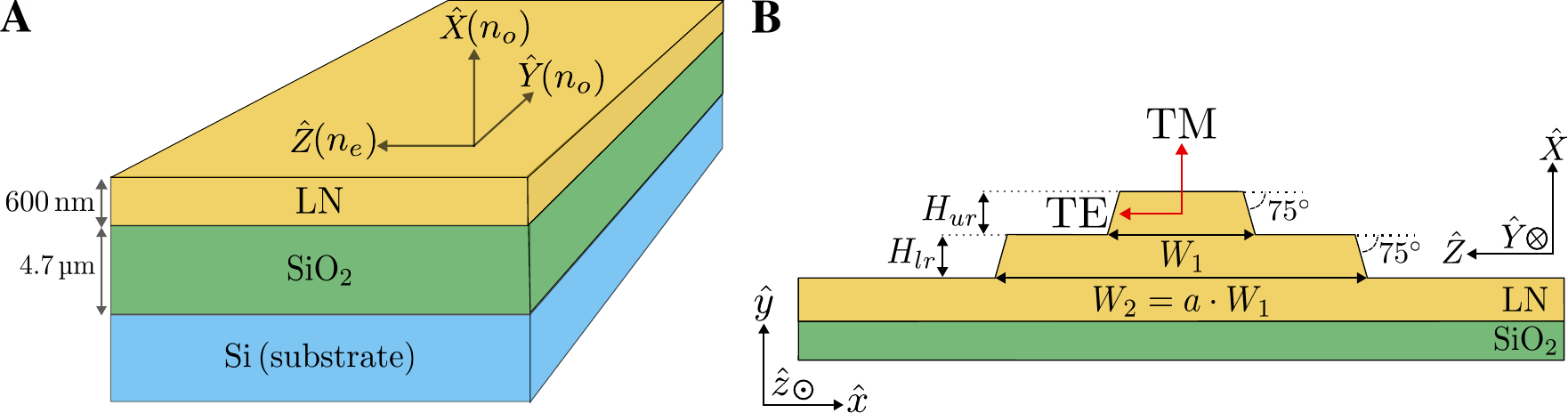}
    \caption{X-cut LNOI platform used for the waveguide design. \textbf{(A)} Schematic of the wafer where $\hat X,\hat Y,\hat Z$ are the crystal axes for LN, while $n_o$ and $n_e$ are the ordinary and extraordinary refractive indices, respectively. \textbf{(B)} Cross section of the double ridge waveguide design, where $\hat x,\hat y, \hat z$ in the lower left corner are the unit vectors representing the geometrical coordinate system.}
    \label{fig:design}
\end{figure}

The waveguide design is based on the LNOI wafer illustrated in Figure\;\ref{fig:design}A. The wafer consists of a 600 nm layer of X-cut lithium niobate (LN) grown on a $4.7\,\SI{}{\micro\meter}$ 
silica layer fabricated on a silicon substrate. The contrast of the lower refractive index of silica 
(1.1--1.6 in the VIS-UV \citep{malitson_interspecimen_1965} to the higher index in LN 
(2.0--2.4 \citep{zelmon_infrared_1997}, provides high light confinement in the LN layer. The crystal orientation is given by the upper case $\hat X, \hat Y,$ and $\hat Z$ unit vectors, where $\hat Z$  specify the optical axis associated with the extraordinary refractive 
', while the $\hat X$ and $\hat Y$ directions are governed by the ordinary refractive index. The proposed design is a double ridge straight waveguide as illustrated in Figure\;\ref{fig:design}B with $\hat x, \hat y$ and $\hat z$ being unit vectors for the geometrical axes. The waveguide can be fabricated by two separate etching steps with etching depths corresponding to $H_{ur}$ and $H_{lr}$ for the upper and lower ridge, respectively. Both ridges are assumed to have a 75$^{\circ}$ angle to horizontal from etching \citep{kozlov_reactive_2023}. The width, $W_2$, of the lower ridge is scaled to the width, $W_1$, of the upper ridge by the parameter $a$. To fully utilize the birefringence of LN for modal PhM, the propagation direction is chosen to be parallel to the crystal $\hat Y$ axis. Consequently, transverse electric (TE) polarized light is associated with the extraordinary index, while transverse magnetic (TM) polarized light is governed by the ordinary index. Perfect Type-0 PhM is found for $H_{ur} = H_{lr} = \mathrm{200\,nm}$, $W_1=1.21\,\SI{}{\micro \meter}$ and $a=1.3$ between the fundamental pump mode and a higher order signal mode (see Figure\;\ref{fig:modeprofiles}). The primary source of linear loss in dielectric waveguides is scattering events that predominantly originate from side wall roughness \cite{pu_nano-engineered_2018}; \cite[p.93]{hunsperger_integrated_2002}. 
Attenuation coefficients are simulated 
using EMode Photonix
for the pump 
(at $\lambda_{\mathrm{p}}$ = 1550 nm)
and for the signal 
(at $\lambda_{\mathrm{s}}$ = 775 nm).
One finds
$\alpha_{\mathrm{p}}$ = 4.9 dB/m 
and $\alpha_{\mathrm{s}}$ = 7.3 dB/m, 
which is of the same order 
as other reported TF-LNOI ridge waveguides 
\citep{wolf_scattering-loss_2018}; \citep{krasnokutska_ultra-low_2018}. 
The loss coefficients were calculated assuming a correlation length of $80\,\mathrm{nm}$ and $100\,\mathrm{nm}$ for the $\hat x$ and $\hat y$ direction, and a $\mathrm{2\,nm}$ and $\,\mathrm{0.2\,nm}$ standard deviation of the roughness for respectively the $\hat x$ and $\hat y$ direction. The loss coefficients should be taken as rough estimates since the linear loss ultimately depends on the quality of fabrication.


%% file: Sections/method.tex
\section{Coupled-amplitude equations}

\begin{figure}[htbp]
\begin{center}
    \includegraphics[width=\textwidth]{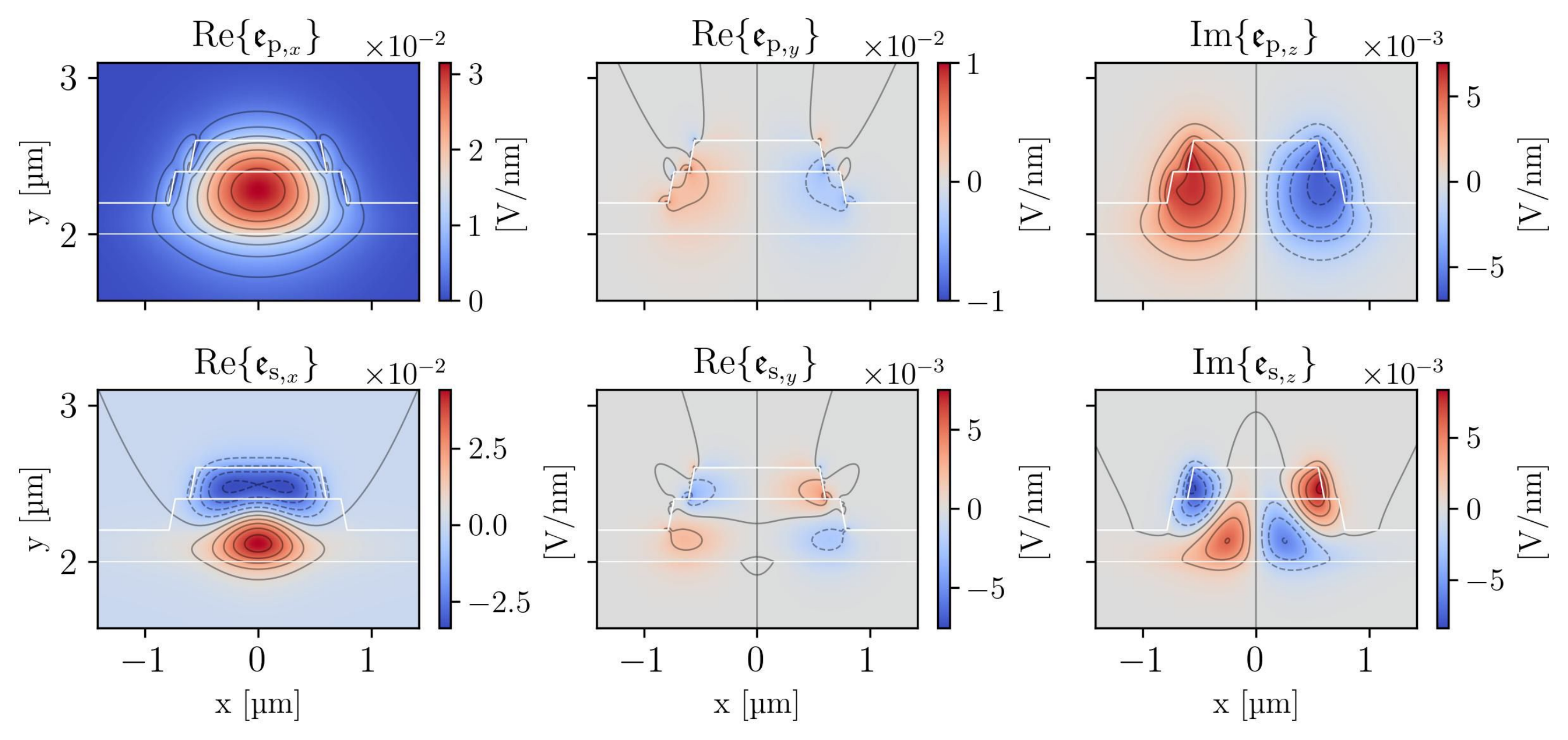}
\end{center}
\caption{Components of the electric mode profile 
for the pump  (upper row) 
and the signal (lower row), which showcase perfect PhM
with the same (quasi-TE) polarization.
The pump mode and the signal mode have equal effective index 
$n \approx 1.896$.
}
\label{fig:modeprofiles}
\end{figure}

The complex wavenumber is expressed as:
\begin{equation}
    k_{\nu} = \beta_{\nu} + i \alpha_{\nu} /2 
    = n_{\nu} \omega_{\nu} / c,
    \quad \nu = \mathrm{p}, \mathrm{s} ,
    \label{eq:k}
\end{equation}
where $n_{\nu}$ is the effective refractive index
of the relevant mode at the angular frequency $\omega_{\nu}$,
and $c$ is the speed of light in vacuum.
The case of continuous-wave (cw)
excitation is considered, 
for which
the complex electric and magnetic fields are 
decomposed as:
\begin{subequations}
\begin{align}
     \vec{\mathcal{E}}_\nu 
    &=  \mathcal{Z}_\nu(z)\vec{\mathfrak{e}}_\nu(x,y)e^{i(k_\nu z-\omega_\nu t)} \\
    \vec{\mathcal{H}}_\nu&=\mathcal{Z}_\nu(z)\vec{\mathfrak{h}}_\nu(x,y)e^{i(k_\nu z-\omega_\nu t )},
\end{align}
\end{subequations}
The 
complex vectors $\vec{ \mathfrak{e}}_{\nu}$ and $\vec{\mathfrak{h}}_{\nu}$ are mode profiles
which are independent of the longitudinal $z$ direction.
The fields are normalized such that: 
\begin{equation}
   N = \frac{1}{2} \iint_{\mathbb{R}^2} \, \left(
   \vec{ \mathfrak{e}}_{\nu}
   \cross \vec{\mathfrak{h}}_{\nu}^* \right)
   \cdot \hat z \, \dif x \dif y 
   = 1\,\mathrm{W}.
\end{equation}
The complex functions $\mathcal{Z}_{\nu}$ are unitless and account for coupling between the pump and signal modes through the coupled-amplitude equations:
\begin{subequations}
\begin{align}
     \partial_z  \mathcal{Z}_\mathrm{p} 
     &= i \frac{\omega_\mathrm{s}}{c}\kappa 
     e^{-\alpha_\mathrm{p}z}e^    {iz\Delta k} \mathcal{Z}_\mathrm{p}^* \mathcal{Z}_\mathrm{s}
    \label{eq:ceq1_SHG} \\
        \partial_z  \mathcal{Z}_\mathrm{s} 
        &= i \frac{\omega_\mathrm{s}}{c}\kappa 
        e^{-iz\Delta k} \mathcal{Z}_\mathrm{p}^2, 
    \label{eq:ceq2_SHG}
\end{align}
\end{subequations}
The power is found from:
\begin{equation}
    P_\nu (z) = N e^{-\alpha_\nu z} \abs{ \mathcal{Z}_\nu(z)}^2 .
    \label{eq:power}
\end{equation}
The complex wavenumber mismatch
is defined as:
\begin{equation}
        \Delta  k =  \Delta \beta +i\Delta \alpha/2 ,
    \label{eq:complmismatch}
\end{equation}
with 
$\Delta \beta = \beta_\mathrm{s} - 2\beta_\mathrm{p}$
and
$\Delta \alpha = \alpha_\mathrm{s}-2\alpha_\mathrm{p}$.
To ensure maximal coupling, 
perfect PhM should be achieved:
\begin{equation}
    \Delta \beta = 0 \quad 
    \Leftrightarrow \quad n_{\mathrm{s}} = n_{\mathrm{p}},
    \label{eq:phmshg}
\end{equation}
 that is, the effective index of the pump must equal that of the signal. 
 To maximize the signal power,
 a large modal overlap is preferred 
 between the pump and the signal.
 The modal overlap is encompassed in the coupling coefficient:
\begin{equation}
    \kappa \equiv \frac{c \epsilon_0}{2 N} \iint_{\mathbb{R}^2} \left(D \vec v\right)\cdot \vec{ \mathfrak{e}}^*_\mathrm{s} \dif x\dif y ,
    \label{eq:kappa}
\end{equation}
with the second-order nonlinear susceptibility tensor $D$
and a vector $\vec{v}$ that contains 
the electric mode profiles. 
Kleinmann's symmetry is assumed, 
and the $D$ tensor for LN  
along with the corresponding $\vec v$ vector 
read \citep{madsen_mid-infrared_2023}:
\begin{equation}
    D = 
    \begin{pmatrix}
        0 & 0 & 0 & 0 & d_{15} & d_{16} \\
        d_{16} & -d_{16} & 0 & d_{15} & 0 & 0 \\
        d_{15} & d_{15} & d_{33} & 0 & 0 & 0 \\
    \end{pmatrix}, \quad \vec v = 
    \begin{pmatrix}
        \mathfrak{e}_{p,_{X}}^2 \\
        \mathfrak{e}_{p,_{Y}}^2 \\
        \mathfrak{e}_{p,_{Z}}^2 \\
        2\mathfrak{e}_{p,_{Y}}\mathfrak{e}_{p,_{Z}} \\
        2\mathfrak{e}_{p,_{X}}\mathfrak{e}_{p,_{Z}} \\
        2\mathfrak{e}_{p,_{X}}\mathfrak{e}_{p,_{Y}} \\
    \end{pmatrix}.
    \label{eq:dLN}
\end{equation}
Notice that the components of $\vec v$ are expressed with respect to the crystal axes. The nonlinear susceptibility tensor contains three unique elements, of which $d_{33} = \mathrm{-27\,pm/V}$ is by far the largest compared to $d_{16} = \mathrm{-2.1\,pm/V}$ and $d_{15} = \mathrm{-4.3\,pm/V}$\citep{zhu_integrated_2021}. From \eqref{eq:kappa} and \eqref{eq:dLN},
one deduces that 
to access $d_{33}$ and thereby achieve highest possible coupling,
both the pump and signal must be TE polarized, \textit{i.e.} 
most of the mode amplitude is along $\hat{x}$.
Mode profiles and effective indices are simulated with EMode Photonix, and perfect PhM is found under such aforementioned conditions between the fundamental pump and the higher-order signal modes displayed in Figure~\ref{fig:modeprofiles}.  
Considering all field components, the coupling coefficient 
(\ref{eq:kappa})
is determined to $\kappa = 2.50\cdot 10^{-5}$.

%% file: Sections/Results.tex
\section{Signal power and conversion efficiency}

\begin{figure}[htbp]
\centering    \includegraphics[width=\textwidth]{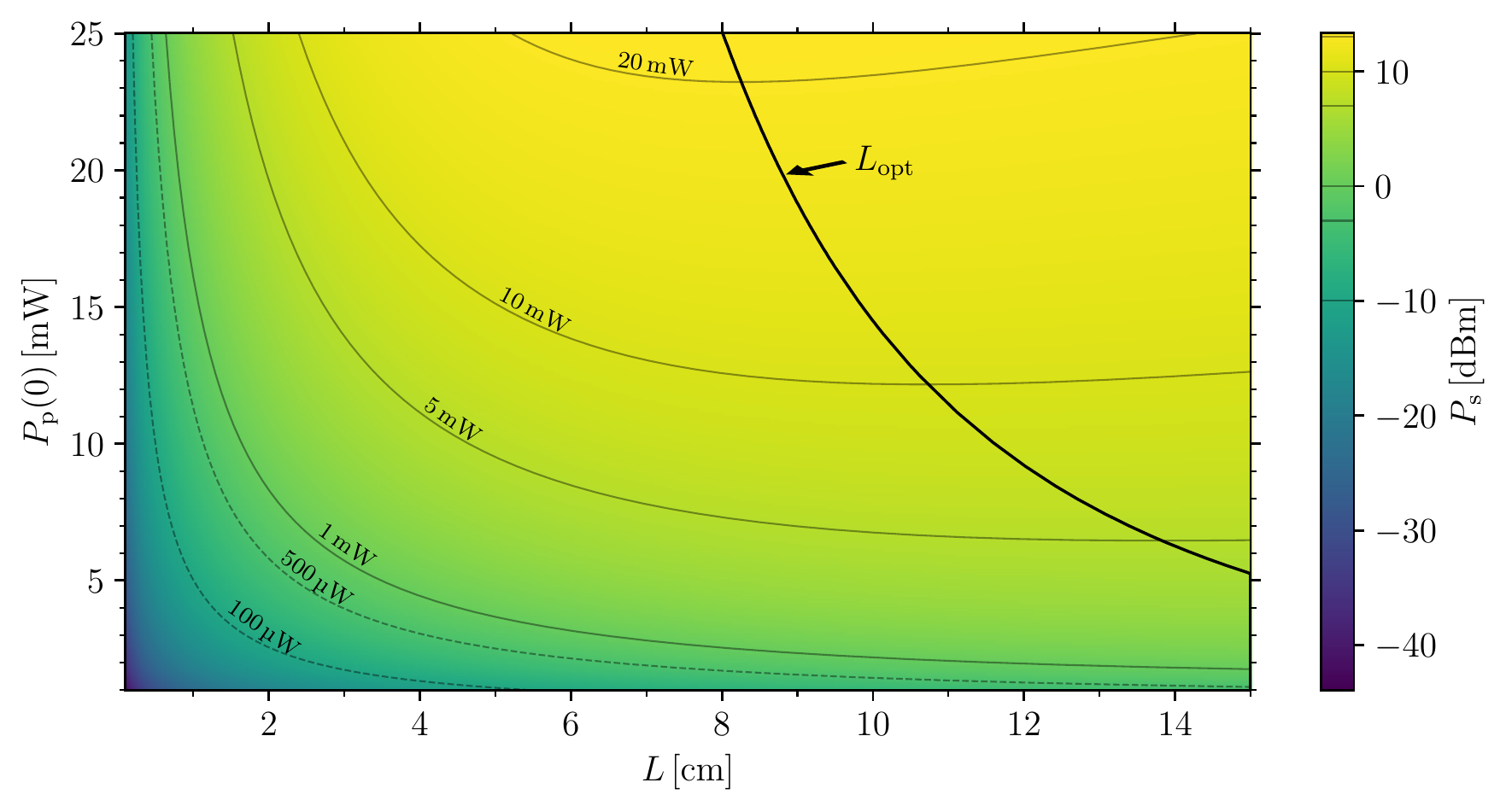}
    \caption{Signal power as a function of interaction length and on-chip pump power for the modes in Figure~\ref{fig:modeprofiles}. Contours are plotted along with the $L_{\mathrm{opt}}$ curve, which marks the optimal interaction length that maximizes the signal power for a given on-chip pump power. Simulation data used for this plot have been linearly interpolated.}
    \label{fig:sigpower}
\end{figure}

The pump power and signal power are computed from \eqref{eq:power} by numerically solving the coupled-amplitude equations \eqref{eq:ceq1_SHG}-\eqref{eq:ceq2_SHG}, for $\mathcal{Z}_p$ and $\mathcal{Z}_s$. The signal power for the modes in Figure~\ref{fig:modeprofiles} is displayed in Figure~\ref{fig:sigpower} as function of the interaction length, $L$, and initial on-chip pump power, $P_\mathrm{p}(0)$. Along with contours, the optimal interaction length, $L_{\mathrm{opt}}$, corresponding to the maximal signal power for a given on-chip pump-power is displayed. Here, it is deduced that 
a waveguide longer than $8\,\mathrm{cm}$ 
is required
to generate a maximal signal power for the considered range of $P_\mathrm{p}(0)$. Straight waveguides longer than a few centimeters are, however, prone to be compromised from fabrication imperfections and is impractical for most on-chip applications. This suggests one to consider spiral single-pass waveguide designs, such that waveguides with lengths $L \sim L_{\mathrm{opt}}$ can be fabricated, while maintaining a small footprint. In this regard, one should consider transitioning to a Z-cut lithium platform with a TM polarized pump and signal, which allows access to the $d_{33}$ element and preserves the PhM condition through bends.

\begin{figure}[htbp]
    \centering
    \includegraphics[width=\textwidth/2]{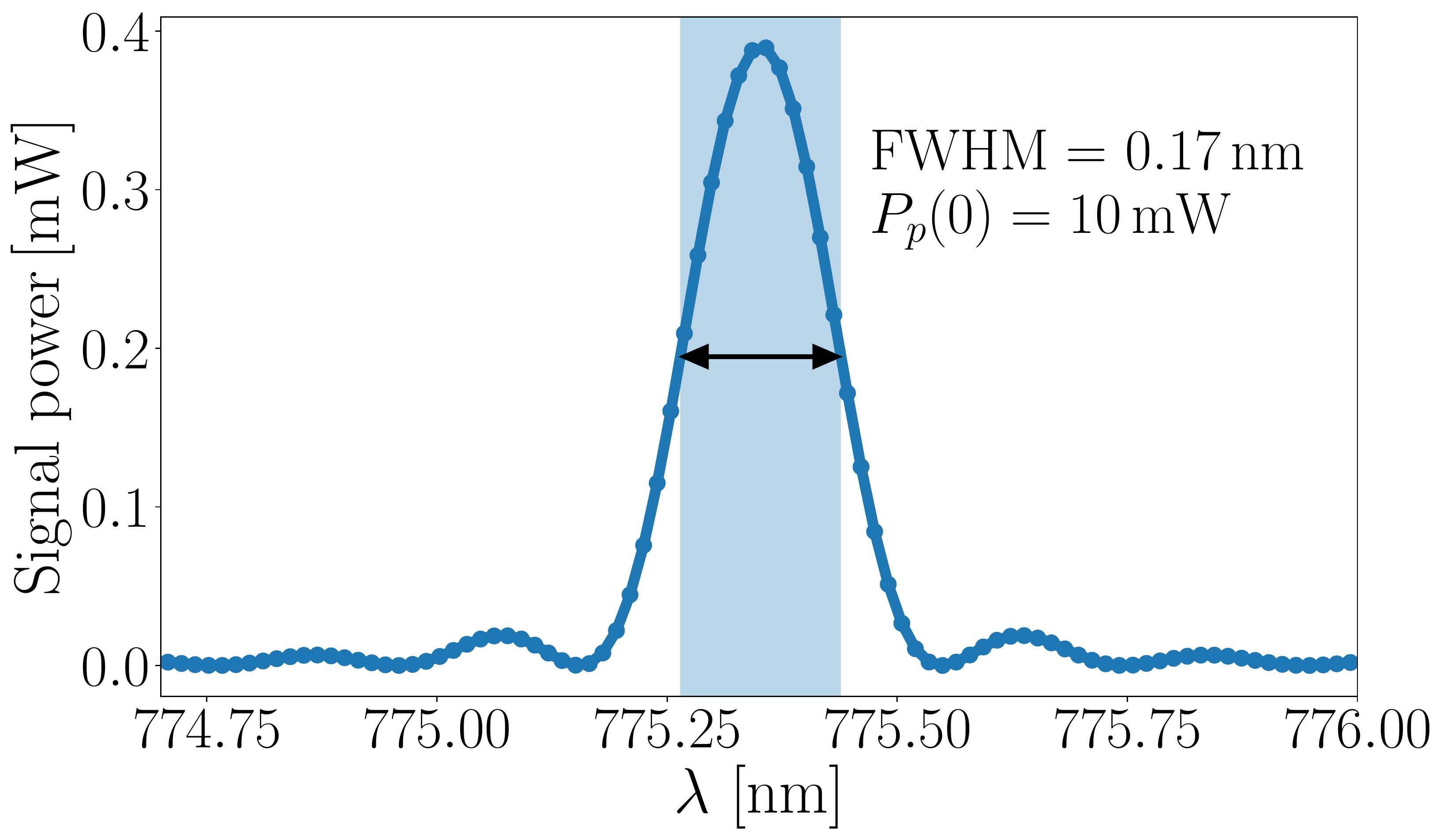}
    \caption{Signal power simulated as a function of the wavelength using  
    an initial on-chip pump power of 10 mW. A FWHM of 0.17 nm is observed, as indicated 
    by the arrow.}
    \label{fig:powersig_fwhm}
\end{figure}

A key metric of interest is the conversion efficiency, which is a measure of the amount of the on-chip pump power, $P_\mathrm{p}(0)$, that is converted to signal power, $P_\mathrm{p}(L)$ after a given interaction length. As evident from Figure~\ref{fig:sigpower}, the conversion efficiency is increasing with $P_\mathrm{p}(0)$ and the interaction length up until $L_{\mathrm{opt}}$, where the conversion efficiency peaks at $86.4\%$ for $P_\mathrm{p}(0)=50\,\mathrm{mW}$ and $L = 8.0\,\mathrm{cm}$.
In the undepleted pump approximation the conversion efficiency is proportional to on-chip pump power \citep{stanton_efficient_2020}, and it is therefore common practice in the literature \citep{hao_second-harmonic_2020};\cite{rao_actively-monitored_2019};\citep{chen_highly-efficient_2018} to define the conversion efficiency per unit pump power: 
\begin{equation}
    \eta = P_{\mathrm{s}}(L) /P_\mathrm{p}^2(0),
    \label{eq:conveff}
\end{equation}
or even further normalizing by the interaction length squared to compensate for the $\eta \propto L^2$ relation. These normalized conversion efficiency definitions are convenient metrics, since they allow one to compare conversion efficiencies for different materials and waveguide designs independently of the waveguide length and on-chip pump power. However, it is important to note that these definitions are based on the undepleted pump approximation, which is only reasonable for short waveguide and low on-chip pump powers. To provide feasible example where the definition in \eqref{eq:conveff} is appropriate, a $1\mathrm{cm}$ waveguide with  $P_\mathrm{p}(0) = 10\,\mathrm{mW}$ is considered. From
Figure~\ref{fig:sigpower} and \eqref{eq:conveff} this yields a conversion efficiency of $\eta = 3.92\,\mathrm{W^{-1}}$.  The corresponding power spectrum of the signal is displayed in 
Figure~\ref{fig:powersig_fwhm} with a full-width-half-maximum (FWHM) of 0.17 nm. 

\subsection{Fabrication tolerances}

\begin{figure}[htbp]
    \centering
    \includegraphics[width=\textwidth]{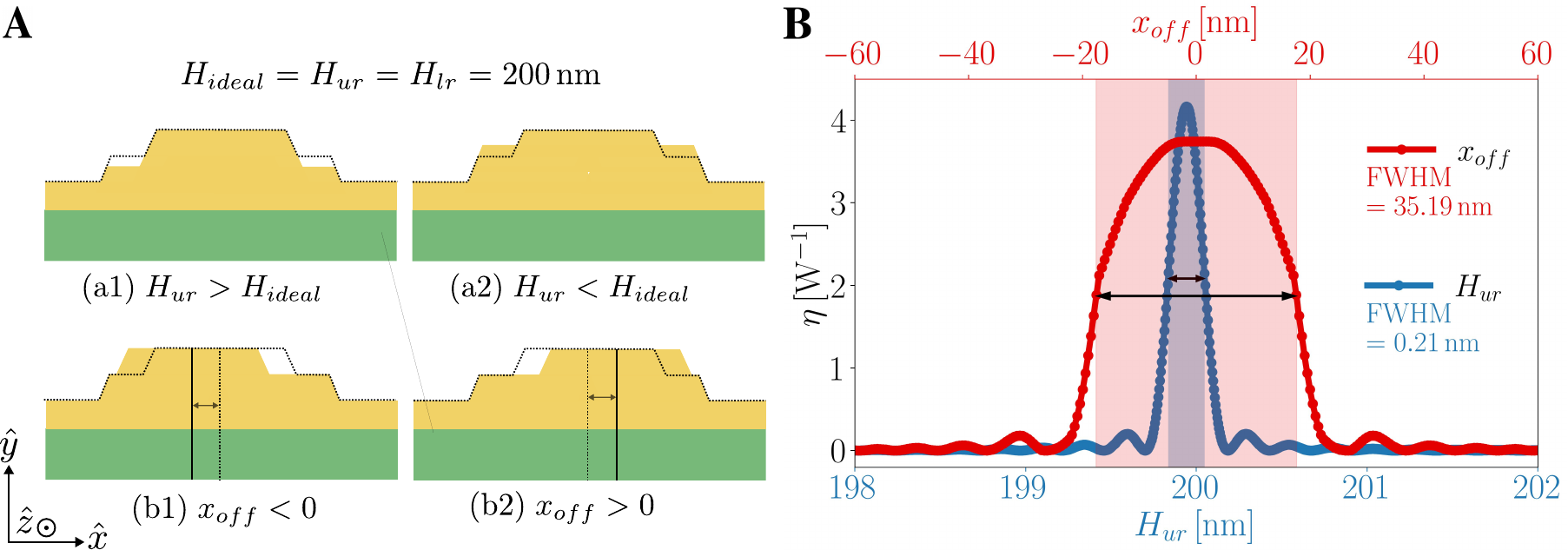}
    \caption{Fabrication tolerances of the conversion efficiency. \textbf{(A)} schematic of the fabrication aberrations from ideal waveguide structure. The contour of the ideal waveguide structure is marked with the dotted line for reference. \textbf{(B)} Conversion efficiency at an $L= \mathrm{1\,cm}$ interaction length as function of the offsets, $x_{off}$, and upper ridge heights, $H_{ur}$. The simulation data used for this plot have been linearly interpolated.}
    \label{fig:tol}
\end{figure}

Deviations from the ideal waveguide geometry are in practice inevitable due to the various uncertainties associated with each intermediate step of waveguide fabrication. To address this, the margins of errors in fabrication are considered, and the impact that these have on the waveguide performance. 
To determine how robust the proposed design is, two parameters that are considered the most likely fabrication aberrations are investigated 
(see Figure~\ref{fig:tol}A). Hence, this is  not an extensive review of the fabrication tolerances of the design. Other aspects could be variations in the sidewall angle or the total thickness of the LN film and the impact that these have on the conversion efficiency \citep{chen_adapted_2023}.

\begin{figure}[htbp]
\begin{center}
    \includegraphics[width=\textwidth]{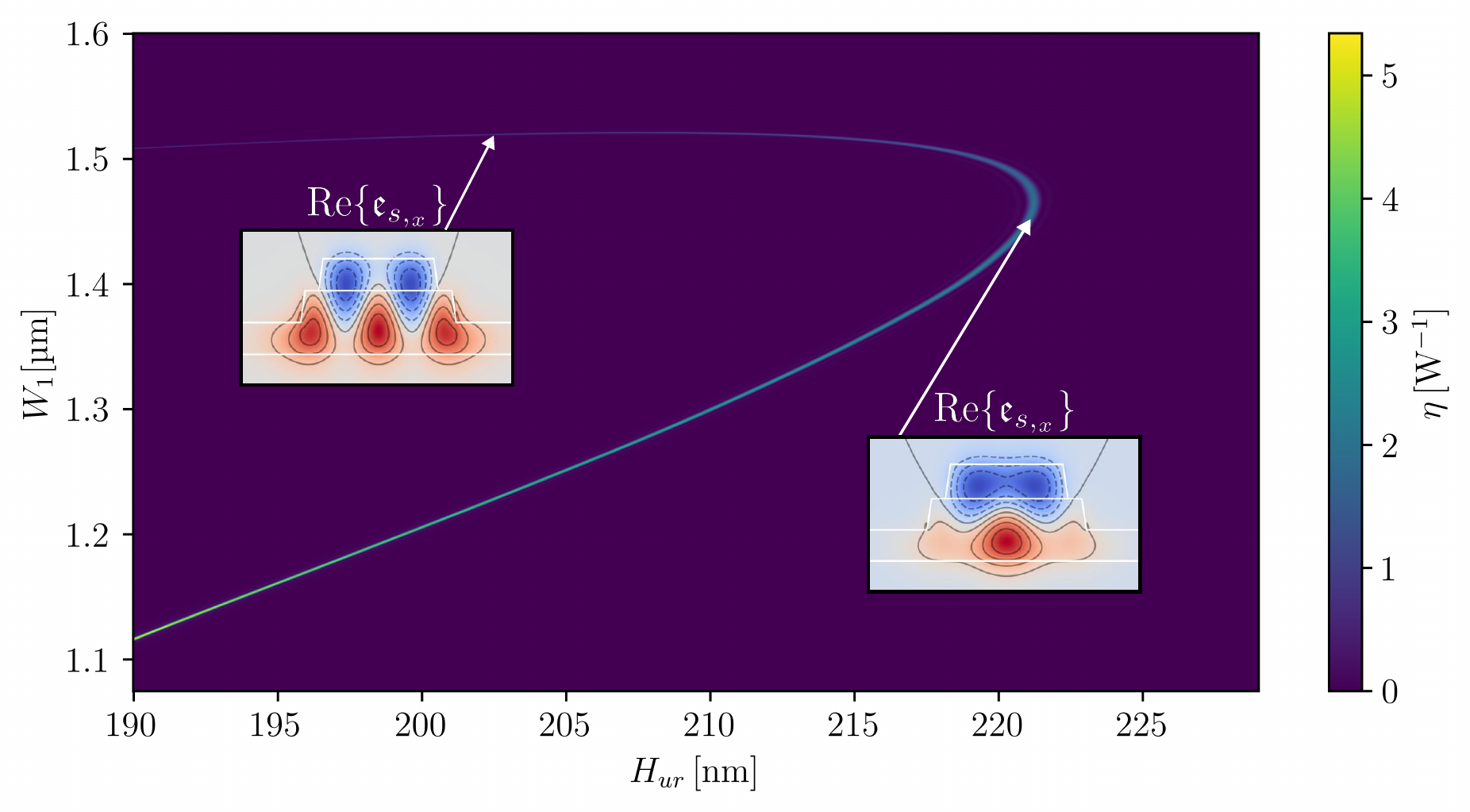}
\end{center}
\caption{Conversion efficiency for the parameter space spanned by $H_{ur}$ and $W_1$ evaluated at a interaction length of $L=1\,\mathrm{cm}$ and scaling parameter $a=1.3$. In addition to the signal mode in Figure\;\ref{fig:modeprofiles} associated with the upward-sloped segment, another higher-order mode displays perfect PhM (top flat segment), though with a lower conversion efficiency. The simulation data used for this plot have been linearly interpolated.
}
\label{fig:conveff_hur_w1}
\end{figure}

The first aberration is the upper ridge height, $H_{ur}$, which depends on how well the etching procedure is calibrated. If the etching rate is higher than expected, the upper ridge height will exceed the target height, i.e.  $H_{ur} < H_{ideal} = \mathrm{200\,nm}$ (a1), while a lower etching rate will yield $H_{ur} < H_{ideal}$ (a2). The second etching step is less critical since the etching time can be adjusted based on the etching rate from the first etching step. The other aberration is misaligned of the upper ridge relative to the lower ridge, defined as the horizontal distance, $x_{off}$, relative to the center of the lower ridge (b1 and b2). While alignment marks are typically placed on the chip to ensure the best possible alignment, misalignment is still a risk that needs to be addressed. Figure~\ref{fig:tol}B displays the impact that the two aberrations have on the conversion efficiency. Here, it is evident that upper ridge height is by far the most critical parameter with a FWHM of $\mathrm{0.21\,nm}$, while the conversion efficiency is more tolerant to misalignment with a FWHM of $\mathrm{35.19\,nm}$. To ensure that a decent amount of signal power can still be generated for $H_{ur} \neq H_{ideal}$, the remaining degrees of freedom for the waveguide geometry parameters are explored. This is done in Figure~\ref{fig:conveff_hur_w1}, where the upper ridge width, $W_1$, is varied to compensate for the non-ideal upper ridge height to obtain PhM and thereby non-vanishing conversion efficiencies.

Here, there is a clear trend, where the conversion efficiency is non-vanishing. The ascending slope segment is associated with the higher-order signal mode in Figure~\ref{fig:modeprofiles}, while the upper flat segment is related to another higher-order TE signal mode. The TE components of the two higher-order signal mode profiles are displayed along with the conversion efficiency trend. The mode profile at $H_{ur}\approx \mathrm{222}\,\mathrm{nm}$ shows where the signal mode in Figure~\ref{fig:modeprofiles} transitions towards the other signal mode at $H_{ur}\approx \mathrm{203}\,\mathrm{nm}$. While the signal mode at $H_{ur}\approx \mathrm{203} \,\mathrm{nm}$ has a lower coupling coefficient, $\kappa = 1.34 \cdot 10^{-5}$, and conversion efficiency, it is more tolerant to variations in $H_{ur}$. By implementing a number of waveguides on a chip with varying $W_1$ in the range 1.1--1.5$\SI{}{\micro \meter}$, one has chip design that is tolerant 
over a range of $\pm 15\,\mathrm{nm}$ from the ideal etching depth.

%% file: Sections/Discussion.tex
\section{Discussion}

The conversion efficiency of $3.92\,\mathrm{W^{-1}}$ at $L=\mathrm{1\,cm}$ is about an order of magnitude lower than the state-of-the-art for single-pass SHG in GaAs-on-insulator \cite{stanton_efficient_2020}. However, GaAs has high absorption at $\mathrm{775\,nm}$ \citep{casey_concentration_1974}, making it an unsuitable choice of material for this particular frequency conversion process. In comparison simulations of similar single-pass LN waveguides  show $\eta = 0.6\,\mathrm{W^{-1}}$ \cite{wang_second_2017} and  $\eta = 0.22\,\mathrm{W^{-1}}$ \citep{luo_highly_2018} at $L = 1\,\mathrm{cm}$, where the presented design has a notably higher conversion efficiency. Substantially higher conversion efficiencies can be achieved in ring resonators \citep{lu_periodically_2019} \citep{boes_efficient_2021}; however, this comes a the cost of a very narrow acceptance band-with  for the phase matching condition. To increase the output power of single-pass design, such as to provide an output power that is viable for practical applications,  multiple arrays of the same waveguides could be combined into a single laser source \citep{madsen_mid-infrared_2023}. However, considering the narrower linewidth of the signal that can be generated from a resonating structure, a ring resonator implementation of the design has prospects for producing spectrally pure photons and, thereby, highly entangled photons. 

The double ridge design was developed as a solution to the design restriction of  
a maximal $\mathrm{200\,nm}$ etching depth. However, recent advances in fabrication have enabled deeper etching, which opens up new waveguide design possibilities. While the higher-order mode associated with the upper flat trend in Figure~\ref{fig:conveff_hur_w1} has a lower conversion efficiency compared to the signal mode in Figure~\ref{fig:modeprofiles}, it is more tolerant to variations in $H_{ur}$. This implies a trade-off between high conversion efficiency and a robust design. Regarding the implementation of the waveguide design on a chip and coupling light into the chip, there is an inherently good overlap between the fundamental TE waveguide mode, Figure~\ref{fig:modeprofiles} (upper row), and a Gaussian pump mode from the fiber, making edge coupling the apparent choice. However, considering the case of SPDC, the fiber mode must couple to the non-Gaussian higher order mode in Figure~\ref{fig:modeprofiles} (lower row). Here, difficulties arise since it is challenging to only couple to a single specific higher-order mode without also exciting other modes \citep{dave_efficient_2019}. 

%% file: Sections/ConclusionOutlook.tex
\section{Conclusion and Outlook}

Using the lithium niobate on insulator platform (Figure~\ref{fig:design}A), a double-ridge waveguide (Figure~\ref{fig:design}B) has been designed for second-harmonic generation at 775 nm from the infrared-C wavelength. Perfect phase matching is achieved 
between the fundamental TE waveguide mode 
at 1550 nm
and a higher-order TE mode at the second harmonic (Figure~\ref{fig:modeprofiles}). This allows access to the largest
element of the second-order nonlinear 
susceptibility tensor. 
From simulations, 
the generated second harmonic has a 
spectral full-width-half-maximum (FWHM) 
of 0.17 nm 
(Figure~\ref{fig:powersig_fwhm}) 
for an initial on-chip pump power of 10 mW.
This makes the design tolerant, and relevant in implementations such as cascaded SPDC \citep{zhang_high-performance_2021}. 
Maximal conversion efficiencies is obtained for waveguides longer than $8\,\text{cm}$, which is beyond what is practicable for integrated circuits fabrication and applications.
Considering a more viable waveguide length of 
1 cm, a conversion efficiency of $3.92\,\mathrm{W^{-1}}$ is obtained, 
which is larger than reported from similar simulation studies of straight single-pass waveguides \cite{wang_second_2017};\citep{luo_highly_2018}. Fabrication tolerances of the proposed waveguide design have been thoroughly investigated. In particular, deviations from the target thickness of the upper ridge have the most critical impact on the conversion efficiency. To compensate for non-ideal etching, the width of the upper ridge could be varied to achieve perfect phase matching and thereby nonvanishing conversion efficiencies. Here, an additional higher-order TE signal mode was found that is robust to variations in upper ridge thickness, though with a lower conversion efficiency. This allows one to find an optimum trade-off between a high-performing and robust design. 

The natural next step is to fabricate and characterize the waveguide performance to validate the numerical results in Figure~\ref{fig:sigpower} and Figure~\ref{fig:powersig_fwhm}. Initially, it should be characterized classically for SHG before advancing to the more complex experimental setup associated with detecting single photons generated from SPDC. In terms of coupling to the higher-order mode for the SPDC application, it makes great sense to optimize a grating coupler design that modulates the pump light such that the modal overlap between the pump and higher-order TE mode is maximized while reducing the amount of power from the pump that couples to the lower-order modes. In addition, there is a potential in using a pulsed laser source, instead of cw considered in this report, to increase the pump power and, thereby, the conversion efficiency. Now, there is a long list of requirements for single photon emitters, hereunder high indistinguishability, temporal purity, brightness, and repetition rates \cite{aharonovich_solid-state_2016}. Before the design qualifies to be a candidate for a single photon source, further work needs to be carried out to investigate the waveguide performance on these parameters in a quantum mechanical theoretical framework. However, as a starting point, the simulation results obtained from the waveguide design indicate an efficient frequency converter with high fabrication tolerances. Moreover, the waveguide design does not require poling, which, in turn, reduces the fabrication complexity.  The design can hence be fabricated with mature technologies, which paves the way for a scalable SPDC source in the infrared-C band.